# Thermal dynamics of charge density wave pinning in ZrTe$_3$

Limin Liu[1,2,*], Changjiang Zhu[1,2,*], Z. Y. Liu[1,3,*], Hanbin Deng[1,2], X. B. Zhou[1,3], Yuan Li[1,2], Yingkai Sun[1,2], Xiong Huang[1], Shuaishuai Li[1], Xin Du[1,2], Zheng Wang[1,2], Tong Guan[1], Hanqing Mao[1], Y. Sui[3], Rui Wu[1,4], Jia-Xin Yin[1,†,‡], J.-G. Cheng[1,2,†], Shuheng H. Pan[1,2,4,5,†]

[1]*Beijing National Laboratory for Condensed Matter Physics, Institute of Physics, Chinese Academy of Sciences, Beijing, 100190, China*

[2]*School of Physical sciences, University of Chinese Academy of Sciences, Beijing, 100049, China*

[3]*School of Physics, Harbin Institute of Technology, Harbin 150001, China*

[4]*Songshan Lake Materials Laboratory, Dongguan, Guangdong 523808, China*

[5]*CAS Center for Excellence in Topological Quantum Computation, University of Chinese Academy of Sciences, Beijing 100190, China.*

[*]Limin Liu, Changjiang Zhu and Z.Y. Liu contributed equally to this work
[†]Corresponding authors: jiaxiny@princeton.edu; jgcheng@iphy.ac.cn; span@iphy.ac.cn
[‡]Current address: Department of Physics, Princeton University, Princeton, NJ 08544, USA

**Impurity pinning has long been discussed to have a profound effect on the dynamics of an incommensurate charge density wave (CDW), which would otherwise slide through the lattice without resistance. Here we visualize the impurity pinning evolution of the CDW in ZrTe$_3$ using the variable temperature scanning tunneling microscopy (STM). At low temperatures, we observe a quasi-1D incommensurate CDW modulation moderately correlated to the impurity positions, indicating a weak impurity pinning. As we raise the sample temperature, the CDW modulation gets progressively weakened and distorted, while the correlation with the impurities becomes stronger. Above the CDW transition temperature, short-range modulations persist with the phase almost all pinned by impurities. The evolution from weak to strong impurity pinning through the CDW transition can be understood as a result of losing phase rigidity.**

In quantum materials, the charge, spin, and orbital of the electronic system can interact and induce collective phenomena, such as superconductivity and CDW [1,2]. CDW is commonly described as a Peierls distortion [1,3], which arises from a combination of Fermi surface nesting and electron-phonon coupling. The real-space modulation of the charge density of a CDW can be expressed as $\rho(x) = \bar{\rho} + \rho_0 \cos(Qx + \varphi)$, where $\bar{\rho}$ is the uniform density, $\rho_0$ is the amplitude of the CDW and $Q=2k_F$ is the wave vector. The phase $\varphi$ describes the location of the CDW relative to the lattice. Frohlich first pointed out that if the wave vector $Q$ is incommensurate with the lattice vector, the energy of the CDW state is independent of the location relative to the lattice [4]. In this case, the CDW can slide through the lattice without resistance. Lee, Rice, and Anderson further pointed out that impurity scattering has a profound impact on the sliding CDW [5]. Specifically, Fukuyama, Lee, and Rice proposed a microscopic model to describe the impurity pinning of a sliding CDW, in which the CDW minimizes its total energy by optimizing both the elastic strain energy due to spatial gradients in $\varphi$ and the pinning energy [6,7]. Two limits that emerge from the model include weak pinning, where the optimum phase is spread over many impurities; and strong pinning, where each impurity is strong enough to pin the CDW phase. It was also pointed out later by Lee, Rice, and Anderson that temperature can introduce strong fluctuations of the CDW order parameter, reducing the global phase rigidity [8]. Inspired by these pioneering theoretical works, we experimentally explore the thermal dynamics of CDW pinning using STM [9]. We observe that in ZrTe$_3$ there is an evolution from weak impurity pinning to strong impurity pinning through the CDW transition, resulting from the temperature tuning of the phase rigidity.

In our experiments, high-quality ZrTe$_3$ single crystals were grown with the iodine chemical-vapor-transport method in a temperature range of 750 to 650 °C. Details about the crystal growth and physical-property characterizations at ambient pressure have been published elsewhere [10].

A home-built low temperature (LT) STM is used for topographic and spectroscopic measurements. Electrochemically etched tungsten tip is used after treatment of field-emission against an Ag film. The samples are mechanically cleaved at a low temperature around 10 K in ultra-high vacuum before insertion into the STM head. All topographic images are acquired with constant current mode. The differential conductance measurements are performed with the standard lock-in technique.

$ZrTe_3$ belongs to the space group $P2_1/m$ with lattice parameters of $a$=0.598 nm, $b$=0.395 nm, $c$=1.043 nm, $\alpha = \gamma = 90°$, $\beta = 97.8°$ [11-13]. As shown in Fig.1(a), it has a layered structure stacked along the $c$-direction. In each layer, 1-D chains along the $b$-direction spread in alignment along the $a$-direction [10,14,15]. Previous diffraction studies found a periodic lattice distortion in $ZrTe_3$ [16-20]. Besides, a Kohn anomaly was observed by phonon measurements [14,16,21,22]. All of these researches have signaled a CDW transition at 70 K. Our resistivity measurements show clear anisotropy, with normal metal behavior along the $b$-direction but with an anomaly at 70 K along the $a$-direction (Fig.1(b)), consistent with the scenario of a CDW transition at 70 K.

To make a direct observation of the CDW, we first acquire the topographic image of the Te terminated surface with atomic resolution. As shown in Fig.1(c), the chain structure of the Te-Te dimers and a few impurities (point defects) can be clearly resolved, with a higher resolution than previous STM results on $ZrTe_3$ [23,24]. The majority of the impurities are likely to be Te vacancies because the vapor pressure of Te is higher than that of Zr and Te can escape from the crystal during the growth process. This speculation is supported by our Energy-dispersive X-ray spectroscopy measurements that show the concentration of Te is about 1% lower than that in the stoichiometric composition, which is consistent with our observation of about 0.5% surface concentration. We note that the local density of states induced by the impurity extend along the $a$-direction that is perpendicular to the atomic chain. Its underlying mechanism requires further investigation with the help of the first-principles calculation, which would be too much beyond the scope of this paper.

Then we perform STS measurements on such atomically flat surfaces. As shown in Fig.1(d), the differential conductance spectrum at 4 K exhibits a 20 meV energy gap around the Fermi level, which is a strong candidate of the CDW gap. With increasing the sample temperature, this gap feature becomes shallower and shallower, and finally diminishes at 70 K. However, at the temperature of 80 K, the overall spectral suppression around the Fermi level persists, indicating the existence of a fluctuating CDW [8,25].

A large-scale STM topographic image with atomic resolution, as shown in Fig. 2(a), clearly displays a quasi-1D modulation along the $a$-direction with a long-range phase coherence characterized by autocorrelation analysis shown in the inset of Fig. 2(a). This observation is consistent with the resistivity anisotropy in Fig. 1(b), supporting that the CDW order along the $a$-direction is developed in the sample. The fast Fourier transformation (FFT) of this topography, as shown in Fig. 2(b), also reveals a high-intensity spot along $a^*$-direction that defines a wave vector of $\sim 0.07a^*$, which agrees with that determined by bulk diffraction measurements [17,18,26].

Furthermore, in the high-resolution topographic images, we observe distortions of the CDW modulation that present the pinning effect of the impurities on the CDW order near the impurities. By careful examination of the lattice, one can also identify the lateral lattice distortions near the impurities. To focus our study on such pinning effects, we extract the CDW intensity map, as shown in Fig. 2(c), by taking an inverse fast Fourier transformation (IFFT) of the FFT areas that refer to CDW (marked by the red dash ellipses in Fig. 2(b)). The black dots represent the impurities. Their locations are extracted from the corresponding high-resolution topographic image. The random distribution of these impurities is demonstrated by the autocorrelation result (Fig. S4). It can be seen that some of the impurities are coinciding with the crest of the CDW modulation. The cross-correlation analysis of the CDW intensity with the positions of the impurities gives a correlation value of 30% (Fig. 2(c) inset), which is close to a weak pinning limit when the optimum phase is spread over many impurities [6,7]. To provide insights into the thermal dynamics of CDW pinning, we systematically study the CDW induced modulation as a function of temperature. Figure 3(a) shows the large-scale topographic images at temperatures ranging from 10 K to 80 K. Although these atomic-resolution large-scale images are not of the exact same microscopic position due to thermal drift as changing temperature, they are still within the same macroscopic area, and therefore they have very similar impurity concentration and distribution. From these images, we find that, as temperature increases, the CDW modulation becomes weaker and the distortion becomes stronger. It is worth noting that at 80 K, a temperature above the CDW phase transition, there are still weak modulations retained forming isolated domains.

To characterize the temperature evolution of the impurity pinning effect on the CDW modulation, we perform FFT analysis of all these topographic images. Along with the temperature increasing, the CDW peaks become lower and broader (shown in Fig. 3(b)), which demonstrates that the intensity of the modulations becomes weaker and less coherent. Using the same analytical scheme for Fig. 2(c), we obtain the CDW intensity map overlayed with the impurity map for all different temperatures, as shown in Fig. 3(c). It can be seen that, as temperature increases, more and more modulation crests locate close to the impurities. Above the CDW transition temperature, the modulation loses its long-range coherence and is almost fully pinned by the impurities, forming small isolated domains. We calculate the cross-correlation between the CDW spatial intensity and the position of the impurities at each temperature and display the results in Fig. 3(d). As a function of distance, the correlation oscillates because of the periodic property of the CDW modulation and its value at zero displacements is the direct measure of the impurity pinning effect on the CDW modulation. Fig. 3(e) shows the value of the correlations without displacement as a function of temperature. As the temperature increases, the correlation progressively increases with a steep increment across the CDW transition, signifying a growing pinning effect with the increasing of temperature. At 80 K, a temperature beyond the CDW phase transition point, the correlation reaches a value of 75%, close to the strong pinning limit, then almost every impurity pins the local phase of CDW [6,7].

With the analysis mentioned above, our experiment, therefore, manifests an intriguing scenario of the thermal destruction of CDW order accompanied by a weak to strong impurity pinning evolution across the CDW transition. The pinning of CDW fundamentally results from the competition between the elastic strain energy due to the spatial gradients of $\varphi$ and the pinning energy [5-7]. The former is related to the phase rigidity of the CDW, which can be progressively destroyed by raising temperature [8]. In 2D CDW material, such as 2H-NbSe$_2$, there is no apparent impurity pinning effect until above the CDW transition temperature [27,28]. The evolution of the phase disordering can be readily visualized in Fig. 4, where we plot the temperature-dependent real-space phase distribution of the CDW modulation. Around the CDW transition temperature, the phase fluctuation increases dramatically, indicating the loss of the long-range phase coherence resulted from the reduction of phase rigidity. This observation, therefore, provides consistent experimental evidence for the theoretical understanding that the thermal dynamics of the weak to strong impurity pinning are closely related to the global phase coherence destruction of the CDW order.

In conclusion, our real-space measurements allow the direct visualization and real-space analysis of the thermal dynamics of the CDW pinning in ZrTe3. Our experimental observations and the results of the analysis are closely related to the pioneering theoretical concepts of impurity pinning at weak and strong limits, as well as fluctuations of the CDW phase. The phase fluctuations emphasized by our experiment can be a generic case for collective ordering phenomena, including superconductivity and various density waves. In the future, it would be interesting to further explore the de-pinning behavior of CDW at different temperatures by applying an in-plane electrical field in STM measurements.

This work was supported by the National Natural Science Foundation of China (Grants No. 11227903, No. 12025408, No. 11834016, and No. 11921004), the Beijing Municipal Science and Technology Commission (Grants No. Z181100004218007 and No. Z191100007219011), the National Basic Research Program of China (Grant No.

2015CB921304), the National Key Research and Development Program of China (Grants No. 2017YFA0302903, No. 2016YFA0302400, and No. 2016YFA0300602), the Strategic Priority Research Program of Chinese Academy of Sciences ( Grants No. XDB07000000, No. XDB28000000, No. XDB33000000, and No. XDB25000000), and the Key Research Program of Frontier Sciences of the Chinese Academy of Sciences (Grant No. QYZDB-SSW-SLH013).

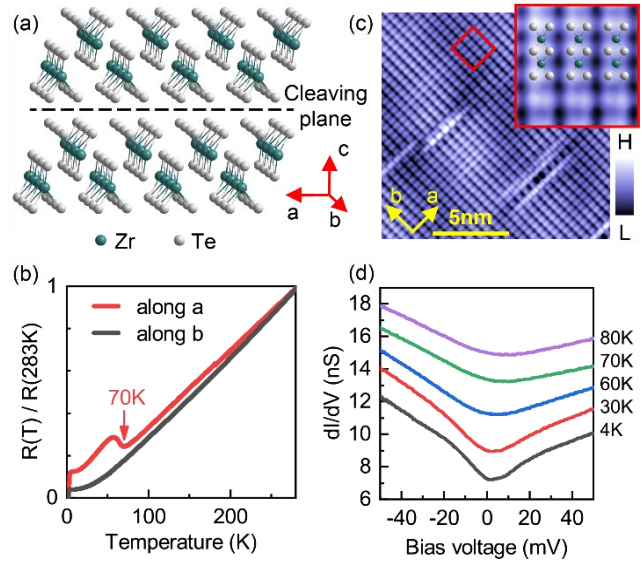

FIG. 1 (a) Crystal structure of ZrTe3. (b) Results of the resistivity measurements along the *a*- and *b*- directions. The anomaly in the resistivity along *a*-direction demonstrates a CDW transition at 70 K. (c) Atomically resolved STM topographic image of the surface of a ZrTe$_3$ single crystal at 4K ($V_b$ = -50 mV, I = 500 pA). inset: the zoomed-in image of a perfect surface area (marked by the red box) with the crystal model laid on top. (d) Tunneling differential conductance at various temperatures (junction setup: $V_b$ = -50 mV, I = 500 pA, and lock-in modulation $V_{mod}$ = 100 uV).

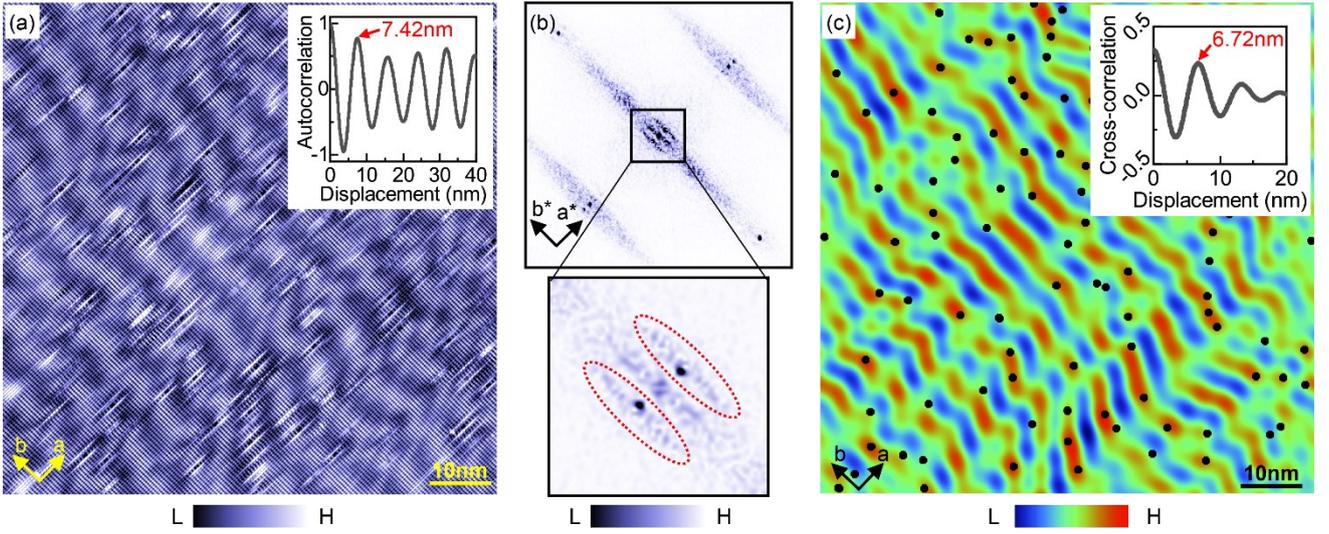

FIG. 2 (a) Large scale topographic image at 4 K ($V_b$ = -50 mV, I = 100 pA), inset is the autocorrelation of the topography with lattice periodicity filtered. (b) FFT of (a), the bottom panel shows the zoomed-in image marked by the black box. (c) The CDW intensity map obtained by taking an inverse fast Fourier transformation (IFFT) of the FFT areas that refer to the CDW modulation (marked by the red dash ellipses in (b)). The black dots in the IFFT map mark the positions of the impurities and the inset shows the cross-correlation of the CDW intensity with the positions of the impurities.

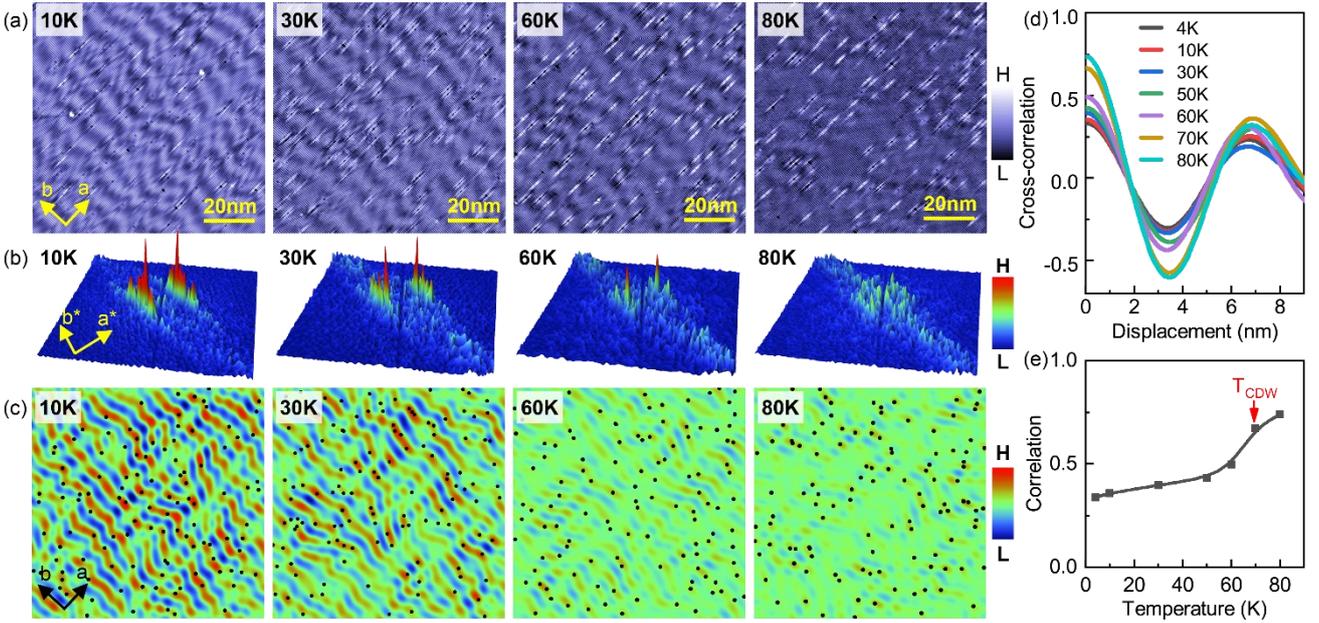

FIG. 3 Temperature evolution of the CDW modulations. (a) STM topographic images at various temperatures. (b) Zoomed-in FFT images of (a) (only the area near the $\bar{\Gamma}$ point). (c) CDW intensity maps at different temperatures with the impurities marked as the black dots. (d) Cross-correlation of the CDW intensity with the impurities at different temperatures. (e) Correlation of the CDW intensity with the impurities as a function of temperature.

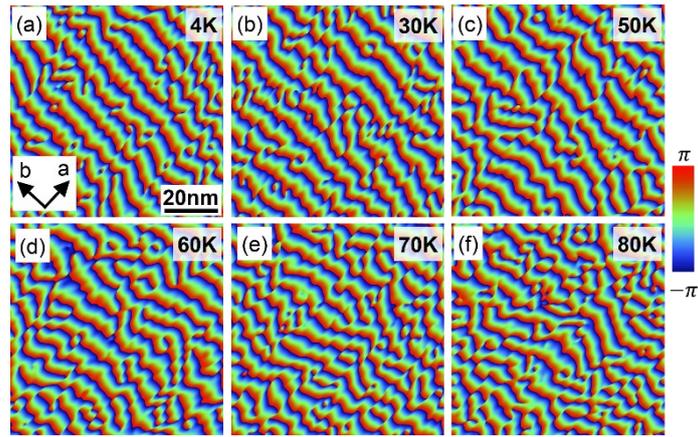

FIG. 4 Real-space phase distribution of the CDW modulation at different temperatures, showing the phase decoherence and fluctuation as increasing temperature. The CDW phase distribution is the phase part of the inverse fast Fourier transformation (IFFT) of the filtered FFT image (See Fig. S6 and notes accompanied).

# Supplementary Information

**Contrast inversion of impurities**

It is noticeable that the impurity states tend to extend along the a-direction that is perpendicular to the atomic chain. The mechanism for such spatial configuration is unclear. We find that it is associated with the electronic structure of the material based on the contrast inversion in the topographic image through inverting the bias voltage as shown in Fig. S1.

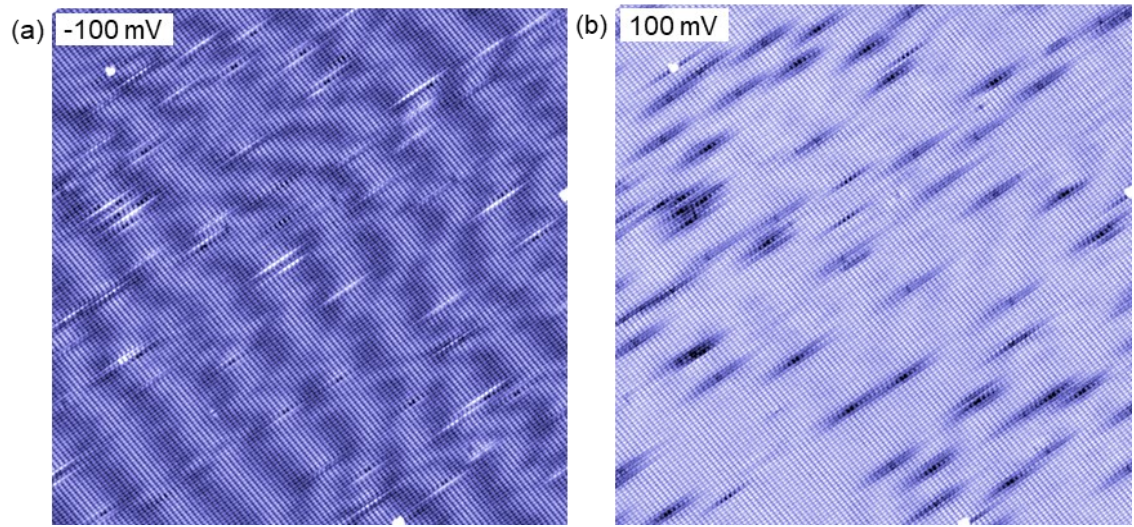

Fig. S1 The STM topographic images at (a) filled state and (b) empty state.

**High-resolution images**

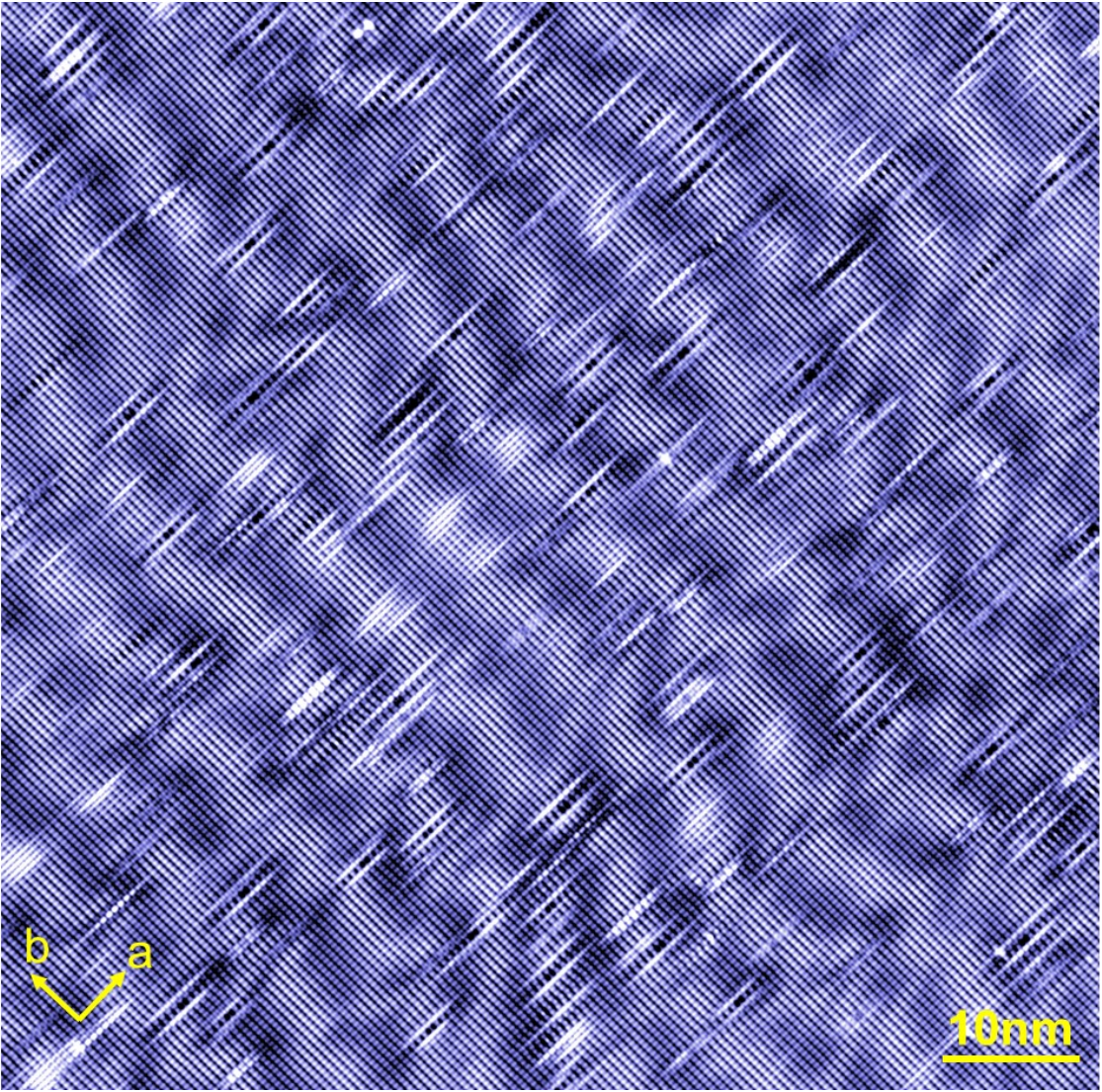

Fig. S2 A high-resolution topographic image at 4K (Fig. 2(a) in the paper).

**The cross-correlation analysis.**

The normalized cross-correlation is processed following the formula below:

$$[f \star g](x,y) = \frac{1}{n}\sum_{x_0,y_0}\frac{[f(x_0,y_0)-u_f][g(x_0+x,y_0+y)-u_g]}{\sigma_f \sigma_g} \quad\cdots\cdots\cdots(1)$$

Here, $f \star g$ is the normalized cross-correlation of f and g; n is the number of pixels of f and g, $u_f$ is the mean of f, $\sigma_f$ is the standard deviation of f, same for $u_g$ and $\sigma_g$.

We first normalize the CDW intensity image and the image of the defect:

$$F(x,y) = \frac{f(x,y)-u_f}{\sqrt{n}\sigma_f}\quad\cdots\cdots\cdots(2)$$

By using formula (1), we then compute the cross-correlation of the normalized CDW intensity map (Fig. S3(a)) and defects distribution (Fig. S3(b)), resulting in a 2D normalized cross-correlation distribution map shown in Fig. S3(c). Since the CDW is in a-direction, the line profile along the a-direction (Fig. S3(d)) shows the correlation of the CDW and the defects, which reaches the maximum at zero displacements, indicating the crest of the CDW intensity is closely related to the position of the defect.

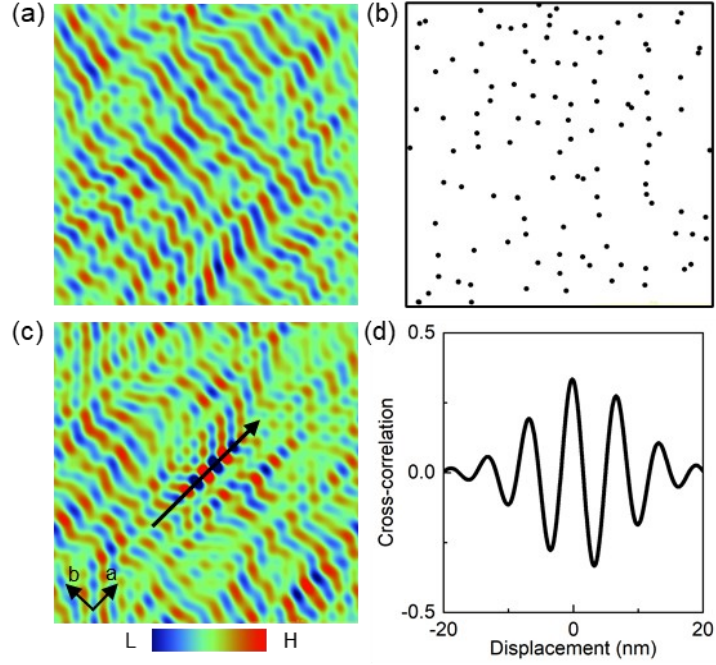

Fig. S3 The process of computing the zero-normalized cross-correlation. (a),(b) Normalized CDW intensity distribution and defects distribution, respectively. (c) The cross-correlation of (a) and (b). (d) A line profile along the black arrow in (c).

**Autocorrelation of the distribution of impurities.**
In 2D materials, the impurities/defects are found to be mobile [1,2]. In ZrTe3, the autocorrelation of the distribution of impurities indicates a random impurities distribution, as shown in Fig. S4.

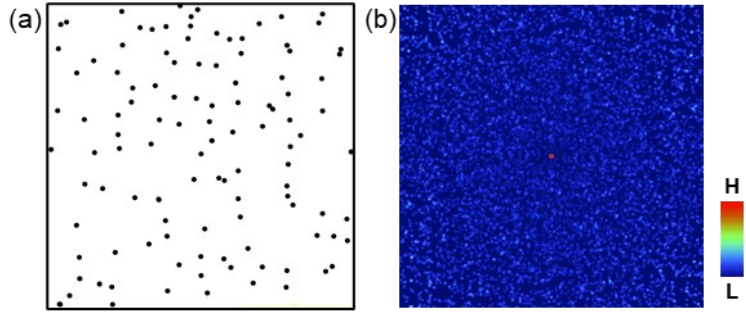

Fig. S4 (a) The distribution of impurities and (b) its autocorrelation, demonstrating a completely random distribution.

**The periodicity variation with temperature**
Recently, a resonant X-ray diffraction study of ZrTe3 has been reported [3]. The authors observed two independent signals which separate at low temperatures and identified a characteristic temperature of 56 K. Our cross-correlation result shows certain consistency with their discovery.  As can be seen in Fig. 3(e). below the temperature of about 56 K, the correlation increases slowly with rising temperature. Above 56 K, the correlation increases dramatically with rising temperature.

   It is noticeable that the periods of autocorrelation and cross-correlation are different.  This difference is primarily due to the pinning effect and the difference between the pristine CDW period and the average distance of the impurities, The period of autocorrelation represents the period of the CDW. In our case, the CDW period at zero pinnings (or very low pinning) is about 7.4 nm and the average distance of the impurities is about 6.8 nm (corresponding to the 0.5% surface concentration). At low temperatures, the pinning effect is very low, so the Autocorrelation period is about 7.4 nm. When temperature increases, the pinning effect gets stronger, the CDW is pinned closer to the impurities. Therefore, the period of the CDW will get closer to the average distance of the impurities. Such effect can be clearly seen in the temperature-dependent autocorrelation and cross-correlation results shown in Fig. S5. It manifests, from another viewpoint, the thermal dynamic evolution of the CDW pinning.

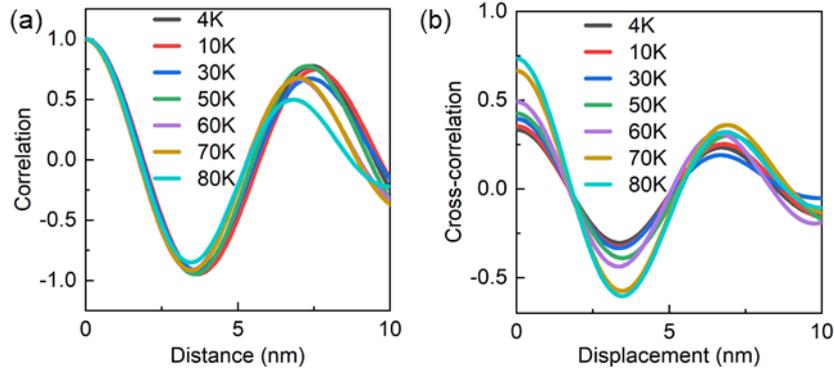

Fig. S5 (a) The autocorrelation and (b) the cross-correlation at different temperatures.

**CDW intensity and phase distribution**

Figure S6 shows the process for obtaining the CDW intensity and phase distributions. We first perform FFT to the original topographic image Fig. S6 (a) to obtain Fig. S6 (b). Then, we filter this data to keep only the area of CDW components, as shown in Fig. S6 (c). An IFFT is performed on the filtered FFT data. The magnitude part [Fig. S6 (d)] and the phase part [Fig. S6 (e)] represents the CDW intensity spatial distribution and phase spatial distribution, respectively.

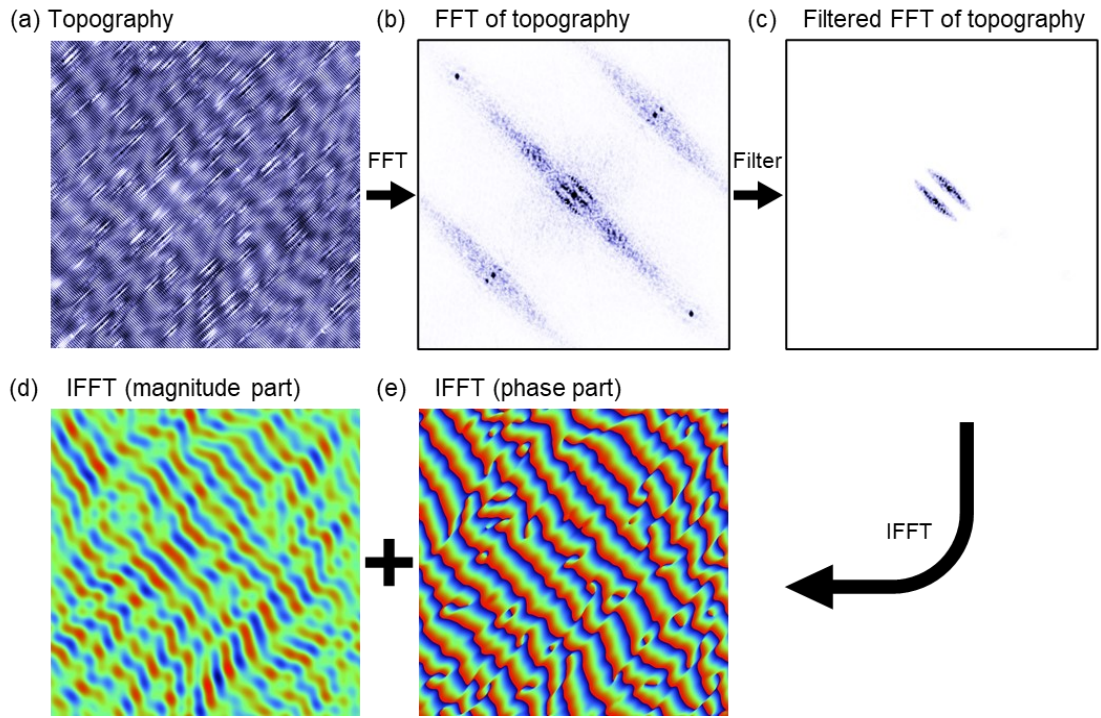

Fig. S6 The process of how the intensity and phase distribution obtained. (a) Original topographic image. (b) FFT of (a). (c) Filtered FFT with only the data around the CDW wave vector left. (d)(e) The magnitude part and the phase

part of the IFFT to (c).